\documentclass[conference]{IEEEtran}
\IEEEoverridecommandlockouts
\usepackage{bbm}
\usepackage{amsmath}
\usepackage{acronym}  
\usepackage[dvips]{color}
\usepackage{epsf}
\usepackage{times}
\usepackage{epsfig}
\usepackage{notoccite}
\usepackage{graphicx}
\usepackage{epstopdf}
\usepackage{pstricks}
\usepackage{amssymb}
\usepackage{amsxtra}
\usepackage{here}
\usepackage{rawfonts}
\usepackage{float}
\usepackage{times}
\usepackage{url}
\usepackage{cite}
\usepackage{caption}
\usepackage{subcaption}
\usepackage{blindtext}
\usepackage{enumitem}
\usepackage{xcolor,cite,etoolbox}
\usepackage{relsize}
\usepackage{lipsum}
\usepackage{graphicx}
\usepackage{tabularx}
\usepackage{xparse}
\usepackage{array}
\newcolumntype{P}[1]{>{\centering\arraybackslash}p{#1}}
\usepackage{mleftright}

\usepackage{mathtools}

\usepackage{graphics}
\usepackage{physics}
\usepackage{amssymb}
\usepackage{siunitx}
\include{newcommands}
\usepackage{multicol}

\usepackage[nomain,acronym,shortcuts]{glossaries}
\makeglossaries
\newcommand*{\acro}[3][]{\newacronym[#1]{#2}{#2}{#3}}

\acro{D2D}{device-to-device}
\acro{SIR}{signal-to-interference-ratio}
\acro{SINR}{signal-to-interference-plus-noise-ratio}
\acro{PCP}{Poisson cluster process}
\acro{CoMP}{coordinated multi-point}
\acro{BS}{base station} 
\acro{MD-CoMP}{macrodiversity CoMP transmission}
\acro{MAC}{medium-access-control}
\acro{JT-CoMP}{joint transmission CoMP}
\acro{CoMP-JT}{coordinated multipoint joint transmission}
\acro{SBS}{small base station}
\acro{MDSD}{multiple devices to a single device}
\acro{MDS}{maximum distance separable}
\acro{SCN}{small cell network}
\acro{PPP}{Poisson point process}
\acro{TCP}{Thomas cluster process}
\acro{CSI}{channel state information}
\acro{PDF}{probability distribution function}
\acro{PMF}{probability mass function}
\acro{RV}{random variable}
\acro{i.i.d.}{independently and identically distributed}
\acro{MBMS}{multimedia broadcasting multicasting service}
\acro{EE}{energy efficiency}
\acro{HCP}{hard-core placement}
\acro{CCDF}{complementary cumulative distribution function}
\acro{CDF}{cumulative distribution function}
\acro{PC}{probabilistic caching}
\acro{RC}{random caching}
\acro{CPF}{caching popular files} 
\acro{PGFL}{probability generating functional}
\acro{KKT}{Karush-Kuhn-Tucker}
\acro{PGF}{point generating function}
\acro{SCA}{successive convex approximation}
\acro{HD}{high-definition}
\acro{FHD}{full-high-definition}
\acro{UHD}{ultra-high-definition}
\acro{VR}{virtual reality}
\acro{AR}{augmented reality}
\acro{5G}{fifth-generation}
\acro{QoS}{quality-of-service}
\acro{QoE}{quality-of-experience}
\acro{IoT}{internet of things}
\acro{MHCPP}{Matern hardcore point process}
\acro{LoS}{line-of-sight}
\acro{NLoS}{non-line-of-sight}
\acro{PSD}{power spectral density}
\acro{MEC}{mobile edge computing}
\acro{E2E}{end-to-end}
\acro{THz}{terahertz}
\acro{CLT}{central limit theorem}
\acro{HQ}{High Quality}
\acro{eMBB}{enhanced mobile broadband}
\acro{URLLC}{ultra reliable low latency communications}
\acro{mmWave}{millimeter wave}
\acro{EVT}{extreme value theory}
\acro{GEV}{generalized extreme value}
\acro{LIS}{large intelligent surface}
\acro{RIS}{reconfigurable intelligent surface}
\acro{RF}{radio frequency}
\acro{UE}{user equipment}
\acro{MIMO}{multiple-input multiple-output}
\acro{EVaR}{entropic value-at-risk}
\acro{DNN}{deep neural network}
\acro{MDP}{Markov decision process}
\acro{RL}{reinforcement learning}
\acro{RNN}{recurrent neural network}
\acro{ANN}{artificial neural networks}
\acro{LSTM}{long short-term memory}
\acro{ReLu}{rectified linear unit}
\acro{VaR}{value-at-risk}
\acro{SNR}{signal-to-noise ratio}
\acro{AoSA}{array of subarray}
\acro{XR}{extended reality}
\acro{AoA}{angle of arrival}
\acro{ULA}{uniform linear array}
\acro{AoD}{angle of departure}
\acro{EM}{electromagnetic}
\acro{HRLLC}{s high-rate and high-reliability low latency communications}
\acro{6DoF}{six degrees of freedom}
\acro{MR}{mixed reality}
\acro{PAPR}{peak to average power ratio}
\acro{OFDM}{orthogonal frequency-division multiplexing}
\acro{OFDMA}{orthogonal frequency-division multiple access}
\acro{SC-FDM}{single carrier frequency-division multiplexing}
\acro{ToA}{time of arrival}
\acro{MUSIC}{multiple signal classification}
\acro{IoE}{Internet of Everything}
\acro{DT}{digital twin}
\acro{PT}{physical twin}
\acro{CT}{cyber twin}
\acro{DRL}{deep reinforcement learning}
\acro{FL}{federated learning}
\acro{DL}{deep learning}
\acro{CRAS}{connected robotics and autonomous system}
\acro{CL}{continual learning}
\acro{MSE}{mean squared error}
\acro{EWC}{elastic weight consolidation}
\acro{ML}{machine learning}
\acro{GD}{gradient descent}
\acro{MLP}{multi layer perceptron}
\acro{TL}{transfer learning}
\acro{UAV}{unmanned aerial vehicle}
\acro{B5G}{beyond 5G}
\acro{CR}{cognitive radio}
\acro{MISO}{multiple-input-single-output}
\acro{STAR-RIS}{simultaneous transmitting and reflecting reconfigurable intelligent surface}
\acro{ES}{energy splitting}
\acro{Tx}{transmitter}
\acro{Rx}{receiver}

\usepackage{datetime}
\usepackage{amssymb}
\usepackage{subcaption}
\usepackage{verbatim}

\usepackage{cite}
\usepackage{amsmath,amssymb,amsfonts}
\usepackage{algorithmic}
\usepackage{graphicx}
\usepackage{textcomp}
\usepackage{xcolor}
\usepackage{mathtools}
\usepackage{gensymb}
\usepackage{optidef}
\usepackage[linesnumbered,ruled,vlined]{algorithm2e}
\SetKwInput{KwInput}{Input}
\SetKwInput{KwOutput}{Output}
\SetKwInput{KwInitialize}{Initialize}
\usepackage{float}

\begin{document}
\title{\huge Simultaneous Transmitting and Reflecting (STAR)-RIS for Harmonious Millimeter Wave Spectrum Sharing
\thanks{This research was supported by the U.S. National Science Foundation under Grants CNS-2030215 and CNS-2030068.}\vspace{-0.4cm}} 
\author{\normalsize Omar Hashash\textsuperscript{1}, Walid Saad\textsuperscript{1}, Mohammadreza F. Imani\textsuperscript{2}, and David R. Smith\textsuperscript{3}  \\ 
\textsuperscript{1}Wireless@VT, Bradley Department of Electrical and Computer Engineering, Virginia Tech, Arlington, VA, USA.\\
\textsuperscript{2}School of Electrical, Computer and Energy Engineering, Arizona State University, Tempe, AZ, USA. \\
\textsuperscript{3}Department of Electrical and Computer Engineering, Duke University, Durham, NC, USA.
\\ Emails:\{omarnh, walids\}@vt.edu, mohammadreza.imani@asu.edu, drsmith@duke.edu
\vspace{-0.45cm}
}
\maketitle
\begin{abstract}
The opening of the millimeter wave (mmWave) spectrum bands for 5G communications has motivated the need for novel spectrum sharing solutions at these high frequencies. 
In fact, reconfigurable intelligent surfaces (RISs) have recently emerged to enable spectrum sharing while enhancing the incumbents' quality-of-service (QoS).
Nonetheless, co-existence over mmWave bands remains persistently challenging due to their unfavorable propagation characteristics.
Hence, initiating \emph{mmWave spectrum sharing} requires the RIS to further assist in improving the QoS over mmWave bands without jeopardizing spectrum sharing demands.
In this paper, a novel \emph{simultaneous transmitting and reflecting RIS (STAR-RIS)}-aided solution to enable mmWave spectrum sharing is proposed.
In particular, the transmitting and reflecting abilities of the STAR-RIS are leveraged to tackle the mmWave spectrum sharing and QoS requirements \emph{separately}.
The STAR-RIS-enabled spectrum sharing problem between a primary network (e.g. a radar transmit-receive pair) and a secondary network is formulated as an optimization problem whose goal is to maximize the downlink sum-rate over a secondary multiple-input-single-output (MISO) network, while limiting interference over a primary network. Moreover, the STAR-RIS response coefficients and beamforming matrix in the secondary network are jointly optimized. To solve this non-convex problem, an alternating iterative algorithm is employed, where the STAR-RIS response coefficients and beamforming matrix are obtained using the successive convex approximation method. Simulation results show that the proposed solution outperforms conventional RIS schemes for mmWave spectrum sharing by achieving a $14.57\%$ spectral efficiency gain.
\end{abstract}
\vspace{-0.2cm}
\section{Introduction}
\vspace{-0.2cm}
The use of \ac{mmWave} frequencies is a cornerstone of current and future wireless cellular systems \cite{semiari2019integrated}. By leveraging the available bandwidth at \ac{mmWave} frequencies, one can enable a plethora of applications ranging from virtual reality to unmanned aerial vehicles~\cite{chen2019artificial}. However, \ac{mmWave} frequencies are already being used to support existing services such as radar systems~\cite{hattab2018toward}. 
As such, enabling mmWave in \ac{B5G} systems requires novel mechanisms for spectrum sharing between incumbent services and new \ac{B5G} applications. Meanwhile, co-existence at these high frequencies faces multiple challenges that range from severe signal attenuation to blockage susceptibility which limit \ac{mmWave} signal propagation~\cite{biswas2021band}.
Henceforth, enabling \emph{\ac{mmWave} spectrum sharing} for B5G applications will require effective schemes that promote \ac{mmWave} signal propagation and fulfill spectrum sharing demands \emph{simultaneously}~\cite{casetti20215g}.

To tackle these requirements, one can deploy \acp{RIS} that have demonstrated a promising potential at \ac{mmWave} bands~\cite{zhang2021millimeter}. In general, \acp{RIS} allow adjustment of the propagation environment by inducing amplitude and phase responses on incident wireless signals.
Despite the abundance of literature on the use of RISs to enhance the \ac{QoS} at \ac{mmWave} bands~\cite{nemati2020ris}, remarkably, only few works look at their use for spectrum sharing between primary and secondary networks~\cite{he2020reconfigurable,yuan2020intelligent,lai2021reconfigurable,wang2020ris,rihan2018optimum}.
For spectrum sharing, the \ac{RIS} can modify the received signals in both primary and secondary networks to control interference levels and enhance spectral efficiency, respectively.
For instance, the work in~\cite{he2020reconfigurable} studies a downlink transmit power minimization problem for an RIS-enhanced cognitive radio (CR) system to enable primary-secondary networks co-existence.
The authors in~\cite{yuan2020intelligent} investigate the deployment of multiple RISs to maximize the data rate over a \ac{MISO}-CR network. In \cite{lai2021reconfigurable}, the authors propose an RIS-enabled successive interference cancellation scheme to enable energy efficient spectrum sharing between transceiver pairs. 
Moreover, the work in~\cite{wang2020ris} uses an RIS to maximize the detection probability of a radar that co-exists with a secondary network, as per the federal communications commission (FCC) recommendations~\cite{rihan2018optimum}. 

Although the works in~\cite{yuan2020intelligent,lai2021reconfigurable,wang2020ris,he2020reconfigurable,rihan2018optimum} showcase diverse applications of \acp{RIS} for spectrum sharing, tailoring such solutions to the \ac{mmWave} regime requires further emphasizing the role of the \ac{RIS} in the interplay between spectrum sharing and propagation enhancement.
Despite its essential application in mitigating potential co-existence interference, the \ac{RIS} plays an indispensable role in overcoming the critical propagation challenges and enhancing the \ac{QoS} in the \ac{mmWave} bands. Hence, the \ac{RIS} is a crucial contributor on both fronts.
Nevertheless, enhancing the \ac{QoS} and mitigating co-existence interference are \emph{tied up to the same \ac{RIS} phase shifts}. Thus, the achieved phase shifts represent a compromise between both disjoint tasks. This can limit the individual performance on each front and prevent the \ac{RIS}-based spectrum sharing scheme from reaching the full potential anticipated at \ac{mmWave} frequencies.
\emph{In essence, enabling \ac{RIS}-based \ac{mmWave} spectrum sharing solutions necessitates decoupling interference mitigation from \ac{QoS} enhancement to explore the full \ac{mmWave} experience in \ac{B5G} networks, an aspect that was not studied in~\cite{yuan2020intelligent,lai2021reconfigurable,wang2020ris,he2020reconfigurable,rihan2018optimum}.}

The main contribution of this paper is a novel \emph{\ac{STAR-RIS}}-aided \ac{mmWave} spectrum sharing solution for 
\ac{B5G} networks. Distinct from conventional reflecting RISs, STAR-RISs can transmit and reflect the incident signals simultaneously, which divides the network into transmitting and reflecting regions accordingly~\cite{liu2021star}. As each of the primary and secondary networks settle in a particular region, the transmitting and reflecting phase shifts can be employed to \emph{separately} limit co-existence interference and enhance the \ac{QoS} at \ac{mmWave} bands.  
In particular, we consider a primary \ac{MISO} network to operate over a licensed \ac{mmWave} band.
In this system, we also have a secondary \ac{MISO} network that comprises \acp{UE} operating under the coverage of a \ac{BS} and function over the same \ac{mmWave} band as that of the primary network. 
To fulfill the demands of both networks, we formulate a spectrum co-existence problem whose goal is to maximize the downlink sum-rate over the secondary network while limiting interference at the primary receiver. As such, the \ac{STAR-RIS} transmission and reflection coefficients along with the beamforming matrix at the secondary \ac{BS} are jointly optimized. To solve this non-convex problem, we use an iterative alternating optimization algorithm that allows us to obtain the \ac{STAR-RIS} response coefficients and beamforming matrix using the \ac{SCA} method. 
\emph{To the best of our knowledge, this is the first work to leverage the transmitting and reflecting abilities of a \ac{STAR-RIS} to separately address the \ac{QoS} and interference requirements of \ac{mmWave} spectrum sharing.}
Simulation results demonstrate the advantages of replacing conventional \acp{RIS} with \acp{STAR-RIS} for mmWave spectrum sharing with a $14.57 \%$ gain in spectral efficiency.

The rest of the paper is organized as follows. Section II presents the system model of the \ac{STAR-RIS}-aided \ac{mmWave} spectrum sharing system. The iterative alternating algorithm used to optimize the \ac{STAR-RIS} coefficients and the beamforming matrix is presented in Section III. Simulation results are presented in Section IV, and conclusions are drawn in Section V.

\vspace{-0.3cm}
\section{System Model}
\subsection{Communication Model}
Consider a \ac{MISO} system composed of a primary \ac{Tx}-\ac{Rx} pair that operate over a licensed mmWave band, as  shown in Fig.~\ref{System Model}. We assume that the \ac{Tx} is equipped with multiple antennas, and it communicates with a single antenna \ac{Rx}.  
In this network, there exists a secondary multi-user \ac{MISO} communication network consisting of a \ac{BS} equipped with $M$ directional antennas that serves a set $\mathcal{K}$ of $K$ single antenna users in the downlink over the same \ac{mmWave} band. To assist the secondary network downlink communication and limit the possible interference that may result on the primary network, a \ac{STAR-RIS} having $N$ equally spaced passive reflecting/transmitting elements is deployed while dividing its surroundings into transmission and reflection regions. Here, we assume that the primary network exists in the transmission region of the \ac{STAR-RIS} while the secondary \acp{UE} exist in the reflective region. In this scenario, the transmitting abilities of the \ac{STAR-RIS} are dedicated to limit interference in the primary network, while the reflecting abilities assist in the secondary downlink communication.
Let $\boldsymbol{h}_{0} \in \mathbb{C}^{M\times 1}$ be the channel gain between the BS and the primary Rx, $\boldsymbol{h}_{k} \in \mathbb{C}^{M\times 1}$ be the channel gain between the BS and each UE $k \in \mathcal{K}$, and $\boldsymbol{H} \in \mathbb{C}^{N\times M}$ be the channel matrix between the BS and the \ac{STAR-RIS}. We also define $\boldsymbol{g}_{0} \in \mathbb{C}^{N\times 1}$ as the channel gain between the \ac{STAR-RIS} and primary \ac{Rx}, and $\boldsymbol{g}_{k} \in \mathbb{C}^{N\times 1}$ as the channel gain between each \ac{STAR-RIS} element $n \in \mathcal{N} \triangleq \{1,\ldots,N\} $ and UE $k$.  
\begin{figure}
\label{System Model}
	\centering
	\includegraphics[scale=0.42]{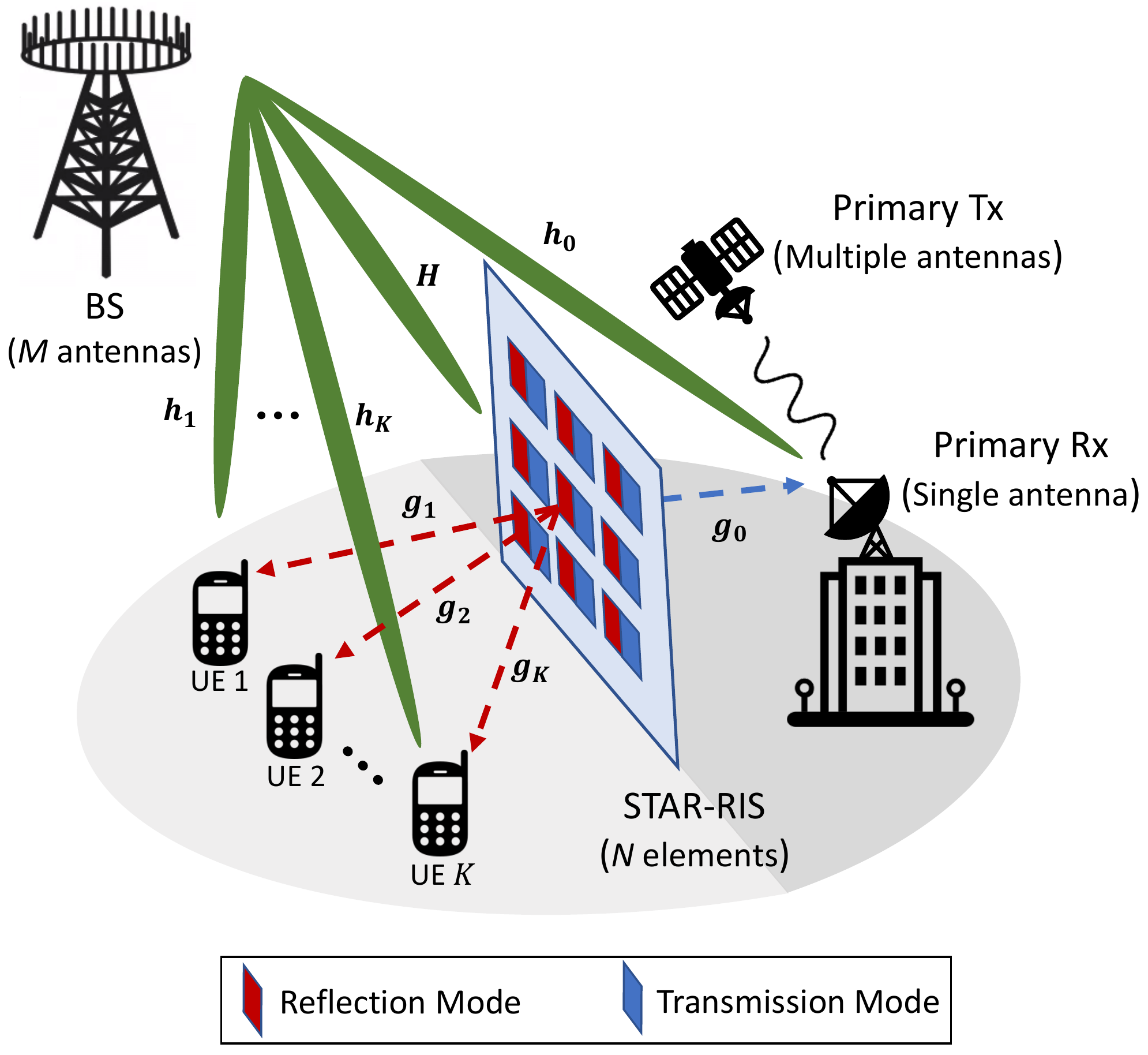}
	\caption{\small{System model of the wireless network with STAR-RIS-aided mmWave spectrum sharing for primary and secondary networks co-existence.}\vspace{-0.5cm}}
	\label{System Model}
\end{figure}
\par In order to initiate downlink communication with multiple UEs, the BS precodes the transmit signal as an $M \times 1$ vector $\boldsymbol{s}= \sum_{k=1}^{K} \boldsymbol{w}_{k}s_{k}$, where $\boldsymbol{w}_{k}\in \mathbb{C}^{M\times 1}$ and $s_k$ are, respectively, the beamforming vector and the unit-power information symbol for each UE $k \in \mathcal{K}$. The power allocation of the BS is subject to a maximum power constraint such that $\lVert \boldsymbol{s} \rVert^2 = \sum_{k \in \mathcal{K}} \lVert \boldsymbol{w}_k \rVert^2 \leq P_{\mathrm{max}}$, where $P_{\mathrm{max}}$ is the maximum transmit power of the \ac{BS}.  

From the STAR-RIS operating modes presented in \cite{mu2021simultaneously}, the \ac{ES} mode is favored as it can simultaneously provide operations over both regions. Hence, to characterize the ES mode for the STAR-RIS, we consider that the energy of the signal incident on each RIS element $n$ is subject to transmission and reflection by assuming that the elements can support electric and magnetic currents~\cite{liu2021star}. Accordingly, the transmission and reflection coefficients of element $n$ can be expressed as $T_n =\sqrt{\beta^t_n}e^{j\theta^t_n}$ and $R_n =\sqrt{\beta^r_n}e^{j\theta^r_n}$ respectively, where $\sqrt{\beta^t_n} \in [0,1], \sqrt{\beta^r_n} \in [0,1]$, $\theta^t_n \in [0,2\pi),$ and $\theta^r_n \in [0,2\pi)$ are, respectively, the amplitude and phase response of the transmission and reflection coefficient of the $n$th element.
The phase responses $\theta^t_n$ and $\theta^r_n$ can be adjusted independently, while the amplitude response of the transmission and reflection coefficients for each element $n$ is governed by the law of conservation of energy such that $|T_n|^2 + |R_n|^2 = 1$~\cite{mu2021simultaneously}. Here, we assume that the passive STAR-RIS elements are ideal with negligible energy consumption. This leads to the following condition:
$\label{energy_conversion}
    \beta^t_n + \beta^r_n = 1.
$
\par The transmission coefficient matrix of the STAR-RIS is given by $\boldsymbol{\Theta}_t$ = diag($T_1, T_2,\ldots,T_N)$.
Similarly, the reflection coefficient matrix is given by $\boldsymbol{\Theta}_r$ = diag($R_1, R_2,\ldots,R_N)$. 
Then, we can define the signal at the primary Rx as
$y = x +(\boldsymbol{h}_0^{H} + \boldsymbol{g}_0^{H} \boldsymbol{\Theta}_t \boldsymbol{H}) \boldsymbol{s} +n_0,$
where $x \in \mathbb{C}$ is the received primary signal and $n_0 \sim \mathcal{C N}(0,\sigma_{0}^{2})$ is the additive white Gaussian noise at the Rx.

In order to effectively separate the propagation environment from the STAR-RIS response, we re-arrange the received signal at the primary Rx equivalently as: 
\vspace{-0.15cm}
\begin{equation}
    y = x +(\boldsymbol{h}_0^{H} + \boldsymbol{\phi}_t^{T} \boldsymbol{G}_0) \boldsymbol{s} +n_0,
    \vspace{-0.1cm}
\end{equation}
where $\boldsymbol{\phi}_t=[T_1, T_2,\ldots, T_n]^{T} \in \mathbb{C}^{N\times 1}$ is the transmission coefficients vector and $\boldsymbol{G}_0$ = diag$(\boldsymbol{g}_0^{H})\boldsymbol{H} \in \mathbb{C}^{N\times M} $ is the cascaded BS-RIS-Rx transmission channel response while removing the effect of the STAR-RIS. The corresponding signal-to-interference-plus-noise (SINR) ratio at the primary Rx will then be: \vspace{-0.2cm}
\begin{equation}
\gamma (\boldsymbol{W}, \boldsymbol{\phi}_t) =  \frac{P_\mathrm{Rx}}{\sum_{k = 1}^{K} |(\boldsymbol{h}_0^H + \boldsymbol{\phi}_t^T \boldsymbol{G}_0) \boldsymbol{w}_{k}|^2 + \sigma_0^2}, 
\end{equation}  
where $P_\mathrm{Rx} = \mathbb{E}[|x|^2]$ is the received primary signal power at the primary Rx and $\boldsymbol{W} = [\boldsymbol{w}_1,\ldots, \boldsymbol{w}_k] \in \mathbb{C}^{M\times K}$is the beamforming matrix at the BS.
Similarly, we define the received signal at UE $k$ as: 
\vspace{-0.1cm}
\begin{equation}
y_k = (\boldsymbol{h}_k^H + \boldsymbol{g}_{k}^H \boldsymbol{\Theta}_r \boldsymbol{H})\boldsymbol{s} + n_k = (\boldsymbol{h}_k^H +  \boldsymbol{\phi}_r^T \boldsymbol{G}_k )\boldsymbol{s} + n_k,
\end{equation}
where $n_k \sim \mathcal{C N}(0,\sigma_{k}^{2})$ is the additive white Gaussian noise at UE $k$, $\boldsymbol{\phi}_r = [R_1, R_2,\ldots, R_n]^T \in \mathbb{C}^{N\times 1}$ is the reflection coefficients vector, and $\boldsymbol{G}_k$ = diag$(\boldsymbol{g}_k^H)\boldsymbol{H}$ is the cascaded BS-RIS-UE reflection channel response while removing the effect of the STAR-RIS.
We assume that no interference occurs from the primary network to the secondary as the primary \ac{Tx}-\ac{Rx} pair provides a point-to-point service and any power received at the UE from the primary network can be considered to be negligible. This assumption is reasonable because the primary \ac{Rx} has a static position and receives a narrow beamwidth signal. Hence, we can then assume that the signal arriving at each UE $k$ is subject to composite channel interference.


As a result, we can define the total sum-rate for all \acp{UE} as:
\vspace{-0.2cm}
\begin{equation}
    r(\boldsymbol{W}, \boldsymbol{\phi}_r) = \sum_{k=1}^{K} B \log_2 (1+\frac{|(\boldsymbol{h}_k^H +  \boldsymbol{\phi}_r^T \boldsymbol{G}_k) \boldsymbol{w}_k|^2}{\sum \limits_{\substack{i=1\\ i\neq k }}^{K} |(\boldsymbol{h}_k^H +  \boldsymbol{\phi}_r^T \boldsymbol{G}_k ) \boldsymbol{w}_i|^2+ \sigma_k^2}), 
    \vspace{-0.3cm}
\end{equation}
where $B$ is the channel bandwidth.
\subsection{Problem Formulation}
\vspace{-0.1cm}
Our main objective is to optimize the beamforming matrix at the BS and the transmission and reflection coefficients of the \ac{STAR-RIS} so as to maximize the sum-rate over the secondary network, while ensuring that the interference experienced by the primary network is maintained below a minimum threshold. It is assumed that perfect channel state information (CSI) are available at the BS. Hence, we can now define our problem:
\vspace{-0.5cm}
\begin{subequations}
\label{opt1}
\begin{IEEEeqnarray}{s,rCl'rCl'rCl}
& \underset{\boldsymbol{W},\boldsymbol{\phi}_t, \boldsymbol{\phi}_r}{\text{max}} &\quad& r(\boldsymbol{W}, \boldsymbol{\phi}_r) \label{obj1}\\
&\text{subject to} && \gamma (\boldsymbol{W}, \boldsymbol{\phi}_t) \geq \gamma_\mathrm{min}\label{c1},\\
&&& \sum_{k=1}^{K} ||\boldsymbol{w}_k||^2 \leq P_{\mathrm{max}}\label{c2},\\
&&& \beta_n^t + \beta_n^r =1 \quad \forall n \in \mathcal{N}\label{c3},\\
&&& \beta_n^t, \beta_n^r \in [0,1] \quad \forall n \in \mathcal{N}\label{c4},\\
&&& \theta^t_n, \theta^r_n \in [0,2\pi) \quad \forall n \in \mathcal{N}\label{c5}, 
\vspace{-0.23cm}
\end{IEEEeqnarray}
\end{subequations}
where $\gamma_{\mathrm{min}}$ is the minimum SINR required for the primary Rx to decode the signal $x$ successfully. The objective function in \eqref{obj1} is the sum-rate of the secondary network users in the downlink. The minimum SINR requirement for decoding at the primary \ac{Rx} is given in \eqref{c1}. The power budget on the BS is captured by \eqref{c2}. The law of conservation of energy that governs the relationship between the amplitude response of the transmission and reflection coefficient is given in \eqref{c3}. This problem showcases the novelty behind integrating a \ac{STAR-RIS} for spectrum sharing, as it can solely provide a simultaneous solution to separately increase spectral efficiency and limit interference through two independent phase shift coefficients. Hence, the \ac{STAR-RIS} provides additional degrees of freedom for spectrum sharing by leveraging both transmitting and reflecting abilities in comparison to conventional \acp{RIS} with single-sided reflection. However, this comes at the expense of compromising the unity amplitude response of conventional \acp{RIS} with reduced amplitude responses of \acp{STAR-RIS}.   
\par The objective function in \eqref{obj1} is non-convex and it is difficult to obtain a global optimal solution. In the next section, we propose an iterative algorithm by using alternating optimization to determine a sub-optimal solution of the beamforming matrix at the \ac{BS} and the \ac{STAR-RIS} transmitting and reflecting coefficients. 
\vspace{-0.1cm}

\section{Alternating Optimization for Beamforming Matrix and STAR-RIS Coefficients \vspace{-0.1cm}}
In this section, the beamforming matrix at the BS and the STAR-RIS transmission and reflection coefficients are jointly optimized to solve problem \eqref{opt1}. We can fix the values of $\boldsymbol{W}$, and $\boldsymbol{\phi}_t$ and $\boldsymbol{\phi}_r$ alternatively, and solve the problem through an iterative procedure. 
\subsection{STAR-RIS Transmission and Reflection Coefficients Optimization}
\vspace{-0.1cm}
For a fixed value of $\boldsymbol{W}$, the problem in \eqref{opt1} reduces to an optimal transmission and reflection coefficients optimization problem as follows:
\vspace{-0.25cm}
\begin{subequations}
\label{opt2}
\begin{IEEEeqnarray}{s,rCl'rCl'rCl}
& \underset{\boldsymbol{\rho},\boldsymbol{\phi}_t, \boldsymbol{\phi}_r}{\text{max}} &\quad& r(\boldsymbol{\phi}_r) =  \sum_{k=1}^{K} B \log_2 (1+\rho_{k}) \label{obj1-1}\\
&\text{subject to} && \rho_k \leq \frac{|(\boldsymbol{h}_k^{H} +  \boldsymbol{\phi}_r^T \boldsymbol{G}_k) \boldsymbol{w}_k|^2}{\sum_{i=1, i\neq k }^{K} |(\boldsymbol{h}_k^H +  \boldsymbol{\phi}_r^T \boldsymbol{G}_k ) \boldsymbol{w}_i|^2+ \sigma_k^2 }\label{c1-1},\\
&&& \sum_{k = 1}^{K} |(\boldsymbol{h}_0^H + \boldsymbol{\phi}_t^T \boldsymbol{G}_0) \boldsymbol{w}_{k}|^2 + \sigma_0^2 \leq \frac{P_\mathrm{Rx}}{\gamma_\mathrm{min}}\label{c2-1},\\
&&& \eqref{c3}, \eqref{c4}, \eqref{c5},
\end{IEEEeqnarray}
\end{subequations}
where $\boldsymbol{\rho} = [\rho_1, \rho_2,\ldots,\rho_K]$ is a slack vector which ensures that constraint \eqref{c1-1} will always hold with equality to the optimal solution. Constraint \eqref{c2-1} guarantees that the interference at the primary Rx is below threshold. We now define $ \Tilde{\boldsymbol{g}}_{kk} =(\boldsymbol{G}_k \boldsymbol{w}_k)^*$, 
$\Tilde{\boldsymbol{g}}_{ki}=(\boldsymbol{G}_k \boldsymbol{w}_i)^*$, and $\Tilde{\boldsymbol{g}}_{0k} =(\boldsymbol{G}_0 \boldsymbol{w}_k)^*$, and, then, \eqref{c1-1} and \eqref{c2-1} can be written as: 
\begin{equation}\label{rho_k}
    \rho_k \leq \frac{|\Tilde{h}_{kk} +  \Tilde{\boldsymbol{g}}_{kk}^H \boldsymbol{\phi}_r|^2}{\sum_{i=1, i\neq k }^{K} |\Tilde{h}_{ki} +   \Tilde{\boldsymbol{g}}_{ki}^H \boldsymbol{\phi}_r|^2+ \sigma_k^2 },
\end{equation}
\begin{equation}\label{interference new}
    \sum_{k = 1}^{K} |\Tilde{h}_{0k} +  \Tilde{\boldsymbol{g}}_{0k}^H \boldsymbol{\phi}_t|^2 + \sigma_0^2 \leq \frac{P_\mathrm{Rx}}{\gamma_\mathrm{min}},
\end{equation}
with $\Tilde{h}_{kk} = \boldsymbol{h}_k^H \boldsymbol{w}_k$, $\Tilde{h}_{ki} = \boldsymbol{h}_k^H \boldsymbol{w}_i$, and $\Tilde{h}_{0k} = \boldsymbol{h}_0^H \boldsymbol{w}_k$. 
We can now transform~\eqref{rho_k} into: 
\vspace{-0.27cm}
\begin{equation}\label{new_rho_k}
    \rho_k (\sum\limits_{\substack{i=1\\ i\neq k }}^{K} |\Tilde{h}_{ki} +   \Tilde{\boldsymbol{g}}_{ki}^H \boldsymbol{\phi}_r|^2+ \sigma_k^2) 
    - |\Tilde{h}_{kk} +  \Tilde{\boldsymbol{g}}_{kk}^H \boldsymbol{\phi}_r|^2 \leq 0.
\end{equation}
\vspace{-0.1cm}
Thus, problem \eqref{opt2} can be reformulated as: \vspace{-0.21cm}
\begin{subequations}
\label{opt3}
\begin{IEEEeqnarray}{s,rCl'rCl'rCl}
& \underset{\boldsymbol{\rho},\boldsymbol{\phi}_t, \boldsymbol{\phi}_r}{\text{max}} &\quad& r(\boldsymbol{\phi}_r) \label{obj1-2}\\

&\text{subject to}  && |[\boldsymbol{\phi}_t]_n|^2 + |[\boldsymbol{\phi}_r]_n|^2 =1  \quad \forall n \in \mathcal{N}\label{c1-2},\\

&&& |[\boldsymbol{\phi}_t]_n|, |[\boldsymbol{\phi}_r]_n| \in [0,1] \quad \forall n \in \mathcal{N}\label{c2-2},\\

&&& \eqref{interference new},\eqref{new_rho_k}.
\end{IEEEeqnarray}
\end{subequations}
To handle the non-convexity in \eqref{c1-2}, we use the penalty method and problem \eqref{opt3} can be written as:
\vspace{-0.18cm}
\begin{subequations}
\label{opt4}
\begin{IEEEeqnarray}{s,rCl'rCl'rCl}
& \underset{\boldsymbol{\rho},\boldsymbol{\phi}_t, \boldsymbol{\phi}_r}{\text{max}} &\quad& 
r(\boldsymbol{\phi}_r)
+C \sum_{n=1}^{N} (|[\boldsymbol{\phi}_t]_n|^2 + |[\boldsymbol{\phi}_r]_n|^2 -  1) \label{obj1-3}\\
&\text{subject to}  && |[\boldsymbol{\phi}_t]_n|^2 + |[\boldsymbol{\phi}_r]_n|^2\leq 1  \quad \forall n \in \mathcal{N}\label{c1-3},\\
&&& |[\boldsymbol{\phi}_t]_n|, |[\boldsymbol{\phi}_r]_n| \in [0,1] \quad \forall n \in \mathcal{N}\label{c2-3},\\
&&& \eqref{interference new}, \eqref{new_rho_k}.
\end{IEEEeqnarray}
\end{subequations}
where $C$ is a large positive constant that enforces $|[\boldsymbol{\phi}_t]_n|^2 + |[\boldsymbol{\phi}_r]_n|^2 = 1$ for the optimal solution of \eqref{opt4}. To handle the non-convex parts in ~\eqref{obj1-3} and \eqref{new_rho_k}, we use the \ac{SCA} method, and the objective in~\eqref{obj1-3} can be approximated as:
\vspace{-0.3cm}
\begin{align}
\label{newobjective}
r(\boldsymbol{\phi}_r)
+\underbrace{  \begin{aligned} &2C \sum_{n=1}^{N} \mathcal{R}(([\boldsymbol{\phi}_t]_n^{(j-1)})^H ([\boldsymbol{\phi}_t]_n - [\boldsymbol{\phi}_t]_n^{(j-1)})
\\
&+([\boldsymbol{\phi}_r]_n^{(j-1)})^H ([\boldsymbol{\phi}_r]_n - [\boldsymbol{\phi}_r]_n^{(j-1)}) ),\end{aligned}}_{f(\boldsymbol{\phi}_t,\boldsymbol{\phi}_r)}
\end{align}
where $f(\boldsymbol{\phi}_t, \boldsymbol{\phi}_r)$ is the first-order Taylor series of $C \sum_{n=1}^{N} (|[\boldsymbol{\phi}_t]_n|^2 + |[\boldsymbol{\phi}_r]_n|^2 -1)$ and the superscript $(j-1)$ represents the value of the variables at iteration $(j-1)$. Likewise, using the \ac{SCA} method, \eqref{new_rho_k} can be expressed as:  
\begin{multline}
\label{convex_new_rho_k}
    \rho_k (\sum \limits_{\substack{i=1\\ i\neq k }}^{K} |\Tilde{h}_{ki} +   \Tilde{\boldsymbol{g}}_{ki}^H \boldsymbol{\phi}_r|^2+ \sigma_k^2)
     -|\Tilde{h}_{kk} +  \Tilde{\boldsymbol{g}}_{kk}^H \boldsymbol{\phi}_r^{(j-1)}|^2
    \\
    -  2\mathcal{R}\left((\Tilde{h}_{kk} + \Tilde{\boldsymbol{g}}_{kk}^H \boldsymbol{\phi}_r^{(j-1)})^{H} \Tilde{\boldsymbol{g}}_{kk}^H (\boldsymbol{\phi}_r - \boldsymbol{\phi}_r^{(j-1)})\right)
    \leq 0,
\end{multline}
where the later part of \eqref{convex_new_rho_k} is the first-order Taylor series evaluated at $\boldsymbol{\phi}_r$ = $\boldsymbol{\phi}_r^{(j-1)}$.
Eventually, the non-convex problem in \eqref{opt4} can be recast as follows:
\begin{subequations}
\label{opt5}
\begin{IEEEeqnarray}{s,rCl'rCl'rCl}
& \underset{\boldsymbol{\rho},\boldsymbol{\phi}_t, \boldsymbol{\phi}_r}{\text{max}} &\quad& 
r(\boldsymbol{\phi}_r)
+f(\boldsymbol{\phi}_t, \boldsymbol{\phi}_r) \label{obj1-4}\\
&\text{subject to}  
&& \eqref{interference new}, \eqref{c1-3}, \eqref{c2-3}, \eqref{convex_new_rho_k}.
\end{IEEEeqnarray}
\end{subequations}
Problem \eqref{opt4} can be solved using the \ac{SCA} method, where the approximated convex problem in \eqref{opt5} is solved at each iteration.

\subsection{Beamforming Optimization at the BS}
Given the transmission coefficients $\boldsymbol{\phi}_t$ and reflection coefficients $\boldsymbol{\phi}_r$, the problem in \eqref{opt1} can be reformulated as:
\begin{subequations}
\label{opt6}
\begin{IEEEeqnarray}{s,rCl'rCl'rCl}
& \underset{\boldsymbol{W},\boldsymbol{\eta}}{\text{max}} &\quad& \sum_{k=1}^{K} B \log_2 (1+\eta_{k}) \label{obj6}\\
&\text{subject to} &&  \eta_k \leq \frac{|\boldsymbol{z}_{rk} \boldsymbol{w}_k|^2}{\sum_{i=1, i\neq k }^{K} |\boldsymbol{z}_{rk} \boldsymbol{w}_i|^2+ \sigma_k^2 } \label{c1-6}\\
&&& \sum_{k = 1}^{K} |\boldsymbol{z}_{t0} \boldsymbol{w}_{k}|^2 + \sigma_0^2 \leq \frac{P_\mathrm{Rx}}{\gamma_\mathrm{min}}\label{c2-6},\\
&&& \sum_{k=1}^{K} ||\boldsymbol{w}_k||^2 \leq P_{\mathrm{max}}\label{c3-6},
\end{IEEEeqnarray}
\end{subequations}
where $\boldsymbol{\eta}= [\eta_1, \eta_2,\ldots,\eta_K]$ is a slack vector which insures that constraint \eqref{c1-6} always holds with equality to the optimal solution, $\boldsymbol{z}_{rk} = \boldsymbol{h}_k^H +  \boldsymbol{\phi}_r^T \boldsymbol{G}_k$, and $\boldsymbol{z}_{t0} = \boldsymbol{h}_0^H + \boldsymbol{\phi}_t^T \boldsymbol{G}_0$. 

To handle the non-convexity in \eqref{c1-6}, we introduce a new variable $\zeta_k$ to decompose \eqref{c1-6} into 
\begin{equation}\label{z_rk}
    |\boldsymbol{z}_{rk} \boldsymbol{w}_k|^2 \geq \zeta_{k} \eta_{k}
\end{equation} 
and 
\begin{equation}\label{z_rk w_i}
    \sum_{i=1, i\neq k }^{K} |\boldsymbol{z}_{rk} \boldsymbol{w}_i|^2+ \sigma_k^2 \leq \zeta_{k}
\end{equation}

Noting that the term $\boldsymbol{z}_{rk} \boldsymbol{w}_k$ in \eqref{z_rk} can be seen as a real number with an arbitrary rotation of the beamforming vector $\boldsymbol{w}_k$~\cite{yang2021energy}, \eqref{z_rk} will be equivalent to $\mathcal{R}(\boldsymbol{z}_{rk} \boldsymbol{w}_k) \geq \sqrt{\zeta_{k} \eta_{k}}$. By replacing the concave function $\sqrt{\zeta_{k} \eta_{k}}$ with the first-order Taylor series, we can rewrite \eqref{z_rk} as: 
\vspace{-0.15cm}
\begin{equation}
\label{z_rk w_k}
\begin{multlined}
    \mathcal{R}(\boldsymbol{z}_{rk} \boldsymbol{w}_k) \geq \sqrt{\zeta_{k}^{(j-1)} \eta_{k}^{(j-1)}} + \frac{1}{2} \sqrt{\frac{\zeta_{k}^{(j-1)}}{\eta_{k}^{(j-1)}}} (\eta_{k} - \eta_{k}^{(j-1)}) \\+ \frac{1}{2} \sqrt{\frac{\eta_{k}^{(j-1)}}{\zeta_{k}^{(j-1)}}} (\zeta_{k} - \zeta_{k}^{(j-1)}).
\end{multlined}   
\end{equation}

Through these approximations, the non-convex problem in \eqref{opt6} can be approximated as the following convex problem:\vspace{-0.27cm}
\begin{subequations}
\label{opt7}
\begin{IEEEeqnarray}{s,rCl'rCl'rCl}
& \underset{\boldsymbol{W},\boldsymbol{\eta}, \boldsymbol{\zeta}}{\text{max}} &\quad& \sum_{k=1}^{K} B \log_2 (1+\eta_{k}) \label{obj7}\\
&\text{subject to} &&  \zeta_k \geq 0 \quad \forall k \in \mathcal{K}\label{c1-7}\\
&&& \eqref{c2-6}, \eqref{c3-6}, \eqref{z_rk w_i}, \eqref{z_rk w_k}  \label{c2-7},
\end{IEEEeqnarray}
\end{subequations}
where $\boldsymbol{\zeta} = [\zeta_1, \zeta_2, \ldots, \zeta_K]$. Hence, the beamforming optimization problem in \eqref{opt6} can be solved using the SCA method, where the approximated convex problem in \eqref{opt7} is solved at each iteration. Hence, we obtain a convex optimization problem whose local optimum solutions can be reached through solving the preceding approximated problems iteratively. This problem can be efficiently solved using convex optimization solvers. The presented solution is detailed in Algorithm 1.

\begin{algorithm}[t]
    \caption{Alternating Optimization using \ac{SCA}}
    \KwInitialize{$\boldsymbol{\phi}_t^{(0)}, \boldsymbol{\phi}_r^{(0)}, \boldsymbol{W}^{(0)}$. Set iterations to $j=1$.}
    \Repeat{the objective \eqref{obj1} converges}{
    Given $\boldsymbol{W}^{(j-1)}$, solve the transmission and reflection coefficients optimization problem in \eqref{opt2} using the SCA method to obtain $\boldsymbol{\phi}_t^{(j)}$ and $\boldsymbol{\phi}_r^{(j)}$. \\
    Given $\boldsymbol{\phi}_t^{(j)}$ and $\boldsymbol{\phi}_r^{(j)}$, solve the beamforming optimization problem in \eqref{opt6} using the SCA method to obtain $\boldsymbol{W}^{(j)}$.\\
    Set $j=j+1$.
} 
\end{algorithm} 

  
    


\section{Simulation Results and Analysis}
For our simulations, we consider a \ac{BS} having $M=16$ directional antennas and $K=4$ secondary \acp{UE}. The location of the \ac{BS} is fixed at (0,25,0), the location of the \ac{STAR-RIS} is considered at (50,0,0), and the location of the primary \ac{Rx} is considered at (60,5,0). We model the pathloss in dB as $PL= 32.4 + 10 c\log_{10}(d) +20 \log_{10}(f)$ ~\cite{docomo2016white}, where $d$ is the distance between the transmitter and the receiver, $f$ is the operating frequency, and $c$ is the pathloss exponent set as $c=2$ for \ac{LoS} and $c=5$ for non-line-of-sight paths~\cite{xiu2021sum}. 
Unless otherwise stated, we consider $f=28$ GHz, $\gamma_{\mathrm{min}} = 20$ dB, $B=1$ MHz, and $\sigma_k^2$ = $\sigma_0^2$ = $-174$ dBm/Hz~\cite{zhang2021millimeter}. Also, we assume that the channels are dominated by \ac{LoS} paths.

We can express the channels between the BS and STAR-RIS as $\boldsymbol{H} = \frac{\alpha}{\sqrt{PL}} \boldsymbol{a}_r(\theta^r, \psi^r)\boldsymbol{a}_t^H(\theta^t, \psi^t)$, where $PL$ is the pathloss between the \ac{BS} and \ac{STAR-RIS}, $\alpha \in \mathbb{C}$ is the complex channel gain, and $\boldsymbol{a}_r \in \mathbb{C}^{N \times 1}$ and $\boldsymbol{a}_t \in \mathbb{C}^{M \times 1}$ are the normalized array response vectors at the \ac{STAR-RIS} and \ac{BS}, respectively\cite{xiu2021sum}. Moreover, $\theta^r$, $\psi^r$ and $\theta^t$, $\psi^t$ are the azimuth/elevation angles of departure/arrival at the \ac{STAR-RIS} and \ac{BS} accordingly.
In addition, we can express the channels from the \ac{BS} to the \ac{UE} as $\boldsymbol{h}_k = \frac{\alpha_k}{\sqrt{PL_k}}  \boldsymbol{a_t}(\theta^t_k, \psi^t_k)$, 
where $PL_k$, $\alpha_k$, $\theta^t_k$, and $\psi^t_k$ are defined similar to those of $\boldsymbol{H}$. The channel $\boldsymbol{g}_k$ between the \ac{STAR-RIS} and \ac{UE} $k$ is defined similar to $\boldsymbol{h}_k$. Then, we can define the cascaded channel $\boldsymbol{G}_k = \frac{\alpha_k'}{\sqrt{PL_k'}}$diag$(\boldsymbol{g}_k^{H})\boldsymbol{H}$, where $PL_k'$ and $\alpha_k'$ are the pathloss and complex gain of the cascaded channel. Similar definitions hold for the primary network for $\boldsymbol{h}_0$, $\boldsymbol{g}_0$, and $\boldsymbol{G}_0$.
We sample the \ac{UE} positions within a circle of radius $r=5$ m from the STAR-RIS. All statistical results are averaged over a large number of independent runs. In our experiments, we compare the proposed \ac{STAR-RIS} scheme with a conventional \ac{RIS} that is modeled as a reflect-only RIS that is deployed adjacent to another transmit-only RIS ~\cite{wu2022resource}. To perform a fair comparison, each RIS is equipped with $N/2$ elements.
\begin{figure}
\begin{subfigure}{.5\textwidth}
	\centering
	\includegraphics[scale=0.5]{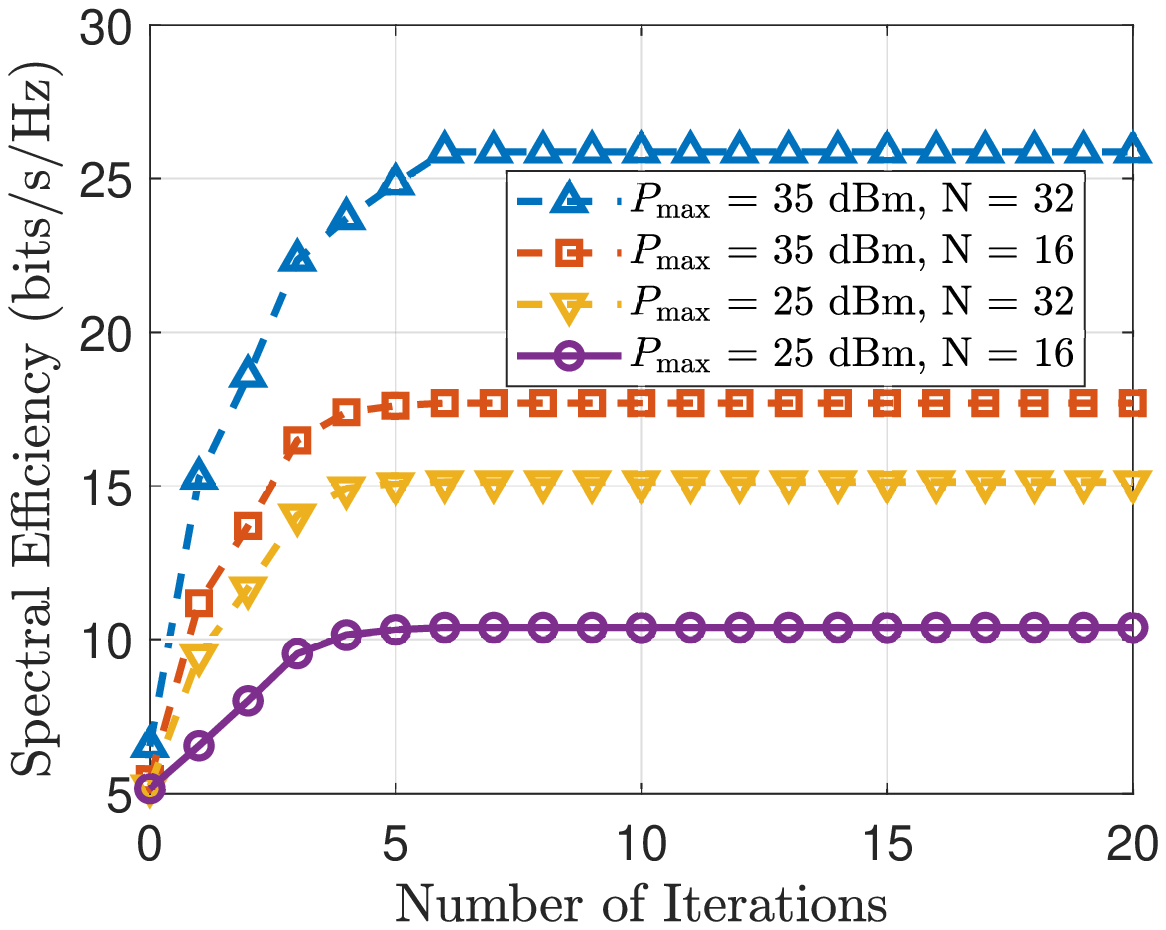}
	\caption{}
	\label{convergence}
\end{subfigure}
\newline
\begin{subfigure}{.5\textwidth}
		\centering
	\includegraphics[scale=0.47]{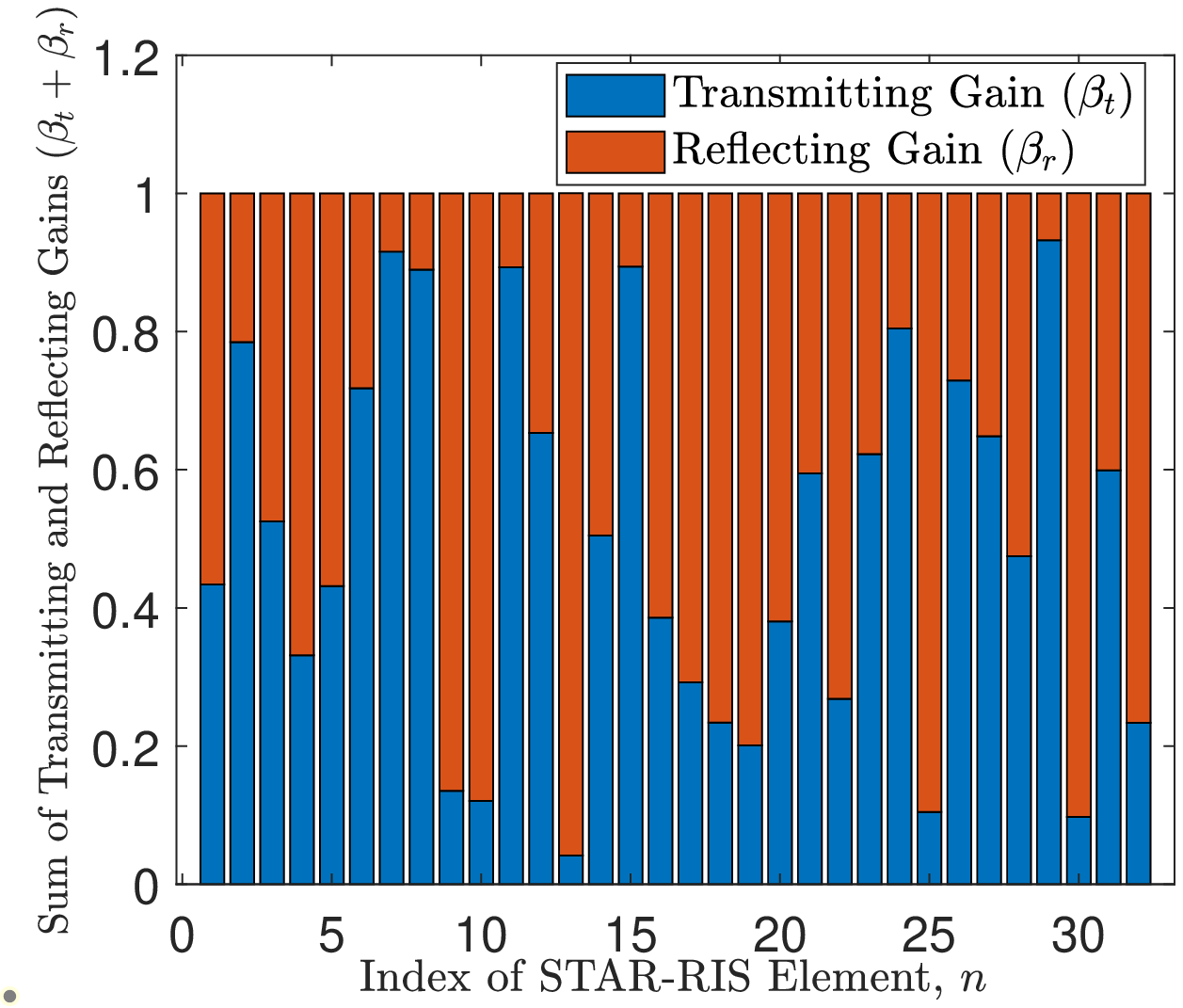}
	\caption{}
	\label{beta_values}
\end{subfigure}
\caption{\small{a) Convergence behavior of the proposed algorithm, b) Sum of transmission and reflection coefficients for $P_{\mathrm{max}} = 35$ dBm and $N =32$ elements.}}
\vspace{-0.5cm}
\end{figure}


\indent Fig.~\ref{convergence} shows the convergence of the proposed \ac{STAR-RIS} scheme for $4$ combinations of simulation parameters over $20$ iterations of the proposed solution. The simulation parameters herein are $P_{\mathrm{max}}= 25$ and $P_{\mathrm{max}} = 35$ dBm, and the number of STAR-RIS elements $N=16$ and $N=32$. From Fig.~\ref{convergence}, we observe that the highest spectrum efficiency recorded was $25.87$ bits/s/Hz that was achieved using $P_{\mathrm{max}} = 35$ dBm and $N=32$ elements. In contrast, a minimal spectral efficiency of 10 bits/s/Hz is reached for $P_{\mathrm{max}} = 25$ dBm and $N=16$ elements. Increasing $P_{\mathrm{max}}$ and $N$ for this minimal scenario by a factor 2, we can see that $P_{\mathrm{max}}$ has a higher impact on the spectral efficiency that increases by around $70\%$ recording $17$ bits/s/Hz. In comparison, increasing $N$ achieves around a $52\%$ spectral efficiency increase that reaches $15.2$ bits/s/Hz. 

\begin{figure}
	\centering
	\includegraphics[scale=0.62]{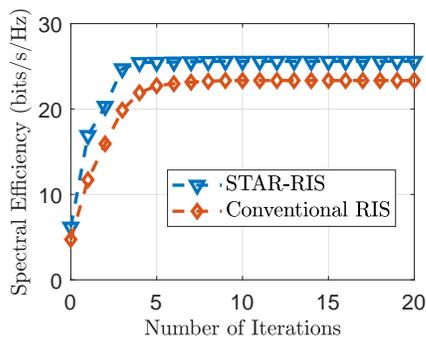}
	\caption{\small{Comparison of the achievable spectral efficiency between the STAR-RIS and the conventional RIS schemes having $P_{\mathrm{max}} = 35$ dBm and $N=32$ elements.}}
	\label{STAR-RIS}
		\vspace{-0.2cm}
\end{figure}
\begin{figure}
	\centering
	\includegraphics[scale=0.7]{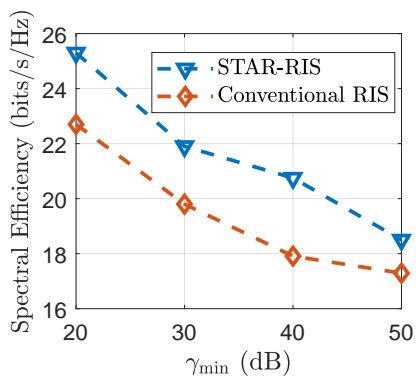}
	\caption{\small{Spectral efficiency achieved by the \ac{STAR-RIS} and the conventional \ac{RIS} schemes with $P_{\mathrm{max}} = 35$ dBm and $N=32$ elements for different $\gamma_{\mathrm{min}}$.}}
	\label{gamma_min}
		\vspace{-0.5cm}
\end{figure}

\indent In Fig.~\ref{beta_values}, we can see that the energy conservation is satisfied for all the indices of the STAR-RIS element. These values are reached after convergence of the proposed algorithm for $P_{\mathrm{max}} =35$ dBm and $N=32$ elements after $20$ iterations. This result showcases how the uniquely dynamic transmission and reflection coefficient magnitudes assist in spectrum sharing, where both amplitudes and phases of the \ac{STAR-RIS} coefficients can contribute to additional degrees of freedom. This is in comparison to reflecting-only \acp{RIS} which neglect the effect of the reflection coefficient magnitudes in spectrum sharing.\\
\indent In Fig.~\ref{STAR-RIS}, we can see that our proposed approach outperforms the conventional RIS scheme with $P_{\mathrm{max}} = 35$ dBm and $N=32$ elements after convergence. In fact, the \ac{STAR-RIS} dominates the conventional \ac{RIS} performance for all iterations. Our proposed scheme achieves a spectral efficiency of $25.87$ bits/s/Hz greater than that of the conventional \ac{RIS} that reaches $22.58$ bits/s/Hz after $20$ iterations. Hence, the proposed scheme achieves a $14.57\%$ spectral efficiency gain over the conventional \ac{RIS} scheme.

\indent In Fig.~\ref{gamma_min}, we can see that our proposed \ac{STAR-RIS} scheme achieves higher spectral efficiency in comparison to the conventional \ac{RIS} scheme for different $\gamma_{\mathrm{min}}$. In particular, the \ac{STAR-RIS} scheme records a maximum of $25.87$ bits/s/Hz in comparison to $22.58$ bits/s/Hz achieved by the conventional \ac{RIS} at $\gamma_{\mathrm{min}} = 20$ dB. 
Clearly, as $\gamma_{\mathrm{min}}$ increases, the amplitude responses become more biased towards the primary network to guarantee the SINR requirement. Consequently, this directly impacts the \ac{QoS} for the secondary \acp{UE}, where the spectral efficiency decreases noticeably.
Although the spectral efficiency of both schemes drop with the increase of $\gamma_{\mathrm{min}}$, the proposed scheme was still able to achieve a better spectral efficiency of around $18.3$ bits/s/Hz higher than that of the conventional \ac{RIS} that reached $17.23$ bits/s/Hz.
\vspace{-0.1cm}
\section{Conclusion}
\vspace{-0.13cm}
In this paper, we have proposed a novel \emph{\ac{STAR-RIS}-aided \ac{mmWave} spectrum sharing} solution to guarantee primary-secondary networks co-existence in B5G systems. In particular, a STAR-RIS topology is leveraged to transmit and reflect the incident signals simultaneously to manage interference and \ac{QoS} demands separately. We have posed the \ac{STAR-RIS} spectrum sharing problem as an optimization problem, where the STAR-RIS response coefficients and beamforming matrix at the \ac{BS} are jointly optimized. Furthermore, the non-convex problem is solved using an alternating iterative algorithm. Numerical results show the superiority of the \ac{STAR-RIS}-aided mmWave spectrum sharing scheme over a conventional \ac{RIS} scheme in terms of spectral efficiency.

\bibliographystyle{IEEEtran}
\def\baselinestretch{0.85}
\bibliography{bibliography}

\end{document}